\documentstyle[epsfig]{aa}

\begin{document}

\thesaurus{11(04.03.1; 11.01.2; 11.19.6; 13.18.1)}


\title{
Optical morphology of distant RATAN-600 radio galaxies
from subarcsecond resolution NOT images 
\thanks{Based on observations made with the Nordic Optical 
Telescope, operated on the island of
La Palma jointly by Denmark, Finland, Iceland, Norway and Sweden, 
in the Spanish Observatorio del Roque de los Muchacos of the
Instituto de Astrophysica de Canarias.} 
}

\author{
T. Pursimo \inst{1}
\and K. Nilsson \inst{1}
\and P. Teerikorpi \inst{1}
\and A. Kopylov \inst{2}
\and N. Soboleva \inst{3}
\and Yu. Parijskij  \inst{2}
\and Yu. Baryshev \inst{4} 
\and O. Verkhodanov \inst{2}
\and A. Temirova  \inst{3}
\and O. Zhelenkova \inst{2}
\and W. Goss \inst{5}
\and A. Sillanp\"a\"a \inst{1}
\and L.O. Takalo \inst{1}
}
\offprints{T.Pursimo}
\institute{
Tuorla Observatory, FIN-21500 Piikki\"o, Finland
\and
Special Astrophysical Observatory, Radio Astronomy Sector,
35147 Niznij Arkhyz, Karachey-Cherkessia, Russia
\and
St.Petersburg Branch of the SAO, Radio Astronomy Laboratory,
Pulkovskoe Shosse 65, 196140 St.Petersburg, Russia
\and
Astronomical Institute of St.Petersburg University, St. Petersburg
198904, Russia
\and
National Radio Astronomical observatory,P.O.Box 0,Socorro,New Mexico
87801,USA  
}

\date{Reveived 27 February 1998 / Accepted 9 September 1998}

\maketitle
\markboth{T.Pursimo et al. NOT imaging of RC/USS radio sources}{}

\begin{abstract}
We present direct imaging data 
of 22 ultra steep spectrum radio sources
obtained at (or near) a subarcsecond seeing.
The basic sample of 40 double  radio sources was selected from 
the RATAN-600 catalogue. 
The FRII-structure has been
confirmed with VLA
and preliminary optical identifications which come from the 6 m-telescope.
As the RATAN-600 flux limit at 3.9 GHz ($\approx$ 10mJy) is fainter
than that of major surveys, the sample may have high-$z$ contents.
This is also suggested by the faint magnitudes in the Hubble diagram.
The final aim is to create a homogeneous sample of high-$z$
radio galaxies in a well defined strip around the sky, with
faint radio limit and subarcsecond morphology down to
$m_{R}=24$.

We could confirm 16 identifications down to $m_{R}\sim$24.
Most of the extended objects have multicomponent structures
as expected from other surveys of high-redshift radio galaxies.
We found five unresolved objects even 
with a subarcsecond seeing. Of the remaining six objects, three 
are extremely faint and the other three have such a complex 
environment that further observations are needed to confirm
the optical identification.

\end{abstract}

\section{Introduction}

There are several reasons to increase the still rather
meagre data on very high-$z$, powerful radio galaxies (e.g.
McCarthy \cite{mcc}, Pariskij et al. \cite{pari:kop}). 
High-$z$ radio galaxies are unique laboratories for investigating the early
stages of galaxy and AGN evolution at look-back times corresponding to 
more than 90\% of the age of the universe derived from
Friedmann models. 
They may be used as tracers of the first generation of galaxy clusters
(Peacock \& Nicholson \cite{pea:nic}; Peacock \cite{peac}) and of the
physical state of the intergalactic space
(Parijskij at al. \cite{pari:goss}). By high-resolution 
one may study important morphological
features related to e.g. 
merging activity and ``star burst'' regions. 

At redshift $z>2$, 120 radio galaxies are known at present
(de Brueck et al. \cite{debr:br}),
in comparison with about 250 radio loud quasars
($S_{5GHz}>$0.03 Jy in Veron-Cetty \& Veron \cite{ver:ver}),
 though the former are intrinsically more abundant. 
According to the popular unified scheme, both classes
are the same thing. One can study the host galaxies and 
close environments of radio galaxies, but this is difficult
for QSOs at a similar redshift.

One aspect, where even a single galaxy may be decisive, is the question of
how close in time to the cosmological singularity it is possible
to find galaxies, with normal stellar population
and supermassive compact objects in their nuclei.
Though the use of high-$z$ objects
in classical cosmological tests is hampered by
severe problems, development of such tests is still
one aim of observational cosmology.
To identify  selection effects and
evolution, large samples are required.
One must increase identifications of very remote galaxies, also
in view of the new generation ground and space telescopes,
which will allow their  study at high resolution.

\subsection{
Extension of identified USS sources to fainter fluxes}

It has been known since the late 70's 
(Tielens et al., \cite{tie:mi}; Blumenthal \& Miley \cite{blu:mi}) 
that radio sources with steep 
spectra are optically fainter (and hence probably more distant) 
than sources with flatter spectra. 
Later it was established 
that observing faint radio sources
with ultra-steep spectra (USS) is an efficient
way  to detect radio galaxies at high redshifts
(see e.g. McCarthy \cite{mcc}).
As the USS Fanaroff-Riley type II 
(FRII-type; Fanaroff \& Riley \cite{fan:ri}) 
radio galaxies are not
good ``standard radio candles''  and as 
the reason  for the success of
the spectrum criterium is not known 
(see e.g. R\"{o}ttgering et al. \cite{rott:lacy}),
it is not clear what the outcome will be when USS samples are extended to 
a progressively fainter flux limit.
Fainter flux may
imply 1) larger redshifts, 2) similar redshifts, though weaker
radio luminosity, or 3) smaller redshifts and still weaker luminosities.
The first alternative is most interesting, though cases 2 and 3 
are also important: extension of the luminosity range will help
one to uncover the influence of radio luminosity on the classical
cosmological tests (angular size--redshift; 
Nilsson et al. \cite{nils:val}
and Hubble diagram; Eales et al. \cite{eal:ra}) and
to decide whether alignment effect depends primarily
on redshift or luminosity.

The flux range where differential normalized source counts 
show steepening is generally regarded as the most promising 
hunting place for high redshift objects.
Parijskij et al. (\cite{pari:bur})  pointed out that the bulk of the
RATAN-600 sample (see below) has fluxes in the range 
of 10-50 mJy at 3.9 GHz where the
normalized counts show a maximum steepening, usually interpreted as
a cosmological effect.
A similar steepening in the counts is seen  separately for steep
spectrum sources (Fig. 6 in Kellermann \& Wall 1987).
It has been suggested (e.g. R\"{o}ttgering et al. \cite{rott:lacy}) 
 that the most effective way to 
find distant galaxies would be a USS sample with 
$S_{408}\sim 0.2-1$ Jy.
Indeed, this has proven to be so
since about 50\% of the R\"{o}ttgering et al. (\cite{rott:lacy})
USS objects have $z>2$ (van Ojik et al. \cite{vanojik:rott}).
The bright end of the USS sources is well studied
(e.g. $4C/USS, B2/1Jy, MRC/1Jy$ McCarthy \cite{mcc}
and references therein) and
recently fainter flux limits have been  reached 
(e.g. $B3/VLA$ $S_{408}>0.8$ Jy Thompson et al. \cite{thompson};
ESO/Key-Project $S_{365}>0.3$ Jy
R\"{o}ttgering at el. \cite{rott:lacy}). 
However, in the  R\"{o}ttgering et al. (\cite{rott:lacy})
sample 365 MHz flux density distribution  peaks at about 1 Jy.

\subsection{RATAN-600 (RC) and UTRAO catalogues}

\begin{figure}
\begin{center}
\hspace*{0.5cm}
\epsfig{file=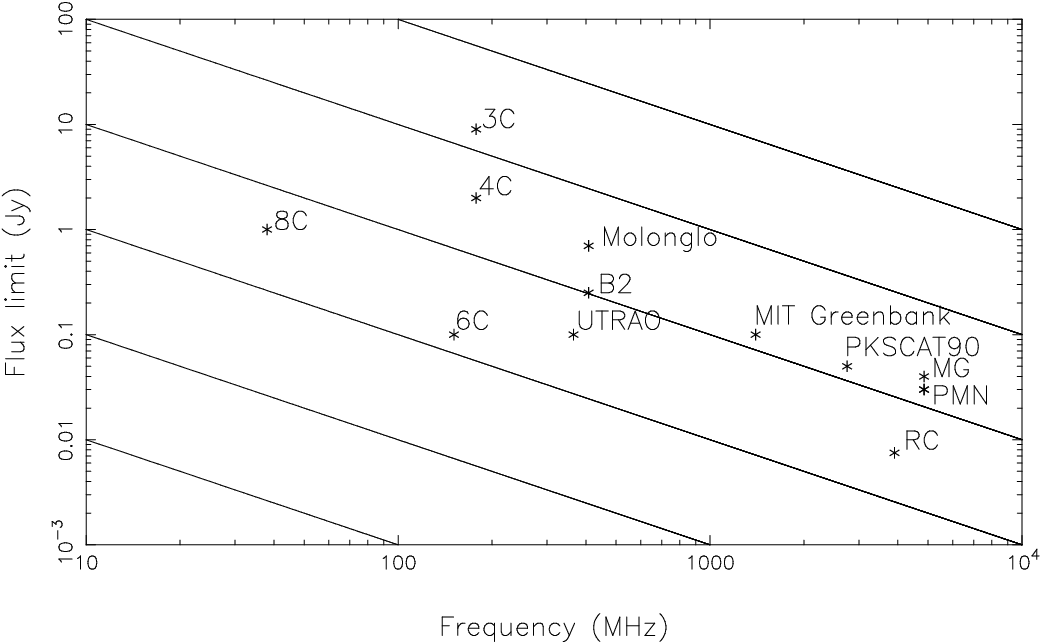, height=5.0cm}
\end{center}
\caption[]{
Frequency - flux limit diagram with the positions of the RC sample
and some other major radio catalogues.  
UTRAO (-36$\degr < \delta <$ 72$\degr$) is the optimum 
low frequency catalogue presently available,  
which can be used for calculating
the spectral index for a large part of the RC sample 
($\delta \sim 5\degr$).
Note that the 6C sample has $\delta >$ 20$\degr$.
The lines correspond to a source with $\alpha$=1.
}
\label{fig1}
\end{figure}

This paper is part of a programme initiated
at the Special Astrophysical Observatory (Russia) with
the aim of searching  distant radio galaxies and
investigating the early evolutionary stages of
the universe (Goss et al. \cite{goss:par}).
We wish to extend the steep-spectrum criteria to fainter fluxes
than previously.
This is accomplished by RC and UTRAO catalogues (see Fig. ~\ref{fig1})

\begin{figure}
\hspace*{0.5cm}
\epsfig{file=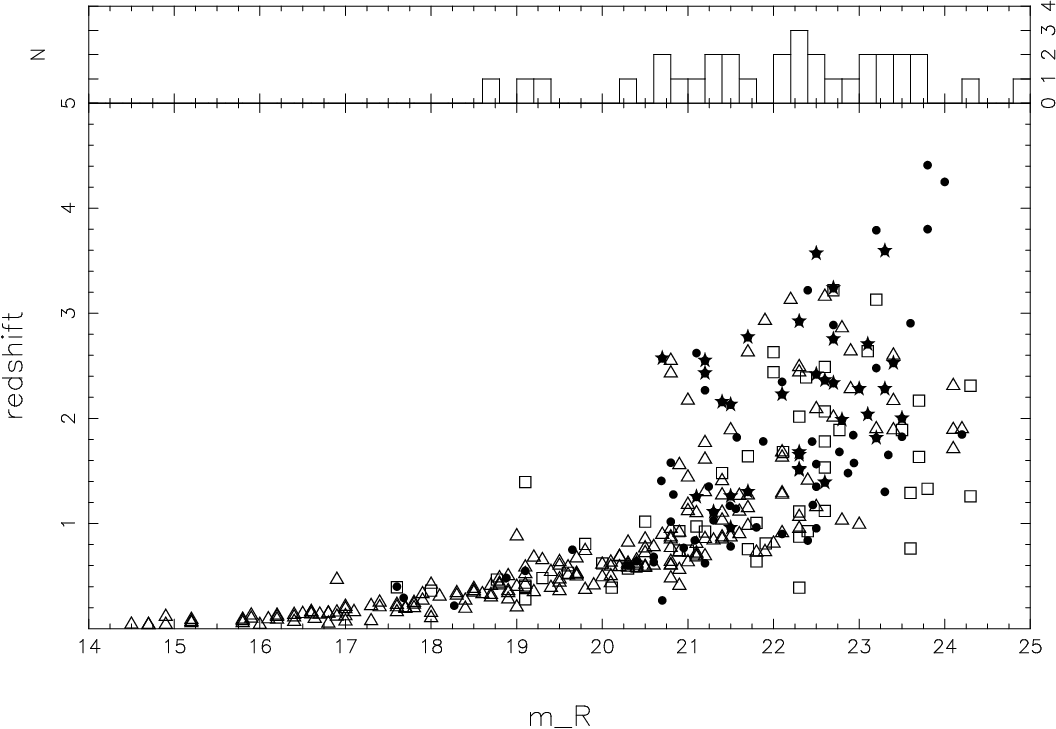, height=5.0cm}
\caption[]{
Hubble diagram in R-band for various radio galaxies 
from the literature.
The triangles are from the  complete Molonglo sample
(McCarthy et al. \cite{mcc:kap2}),
the open boxes are from  
Allington-Smith et al.(\cite{all:spi}),
Maxfield et al. (\cite{max:tho}),
McCarthy et al. (\cite{mcc:spi}),
McCarthy et al. (\cite{mcc:kap1}),
McCarthy et al. (\cite{mcc:van}),
Thompson et al. (\cite{thompson}),
Windhorst et al. (\cite{wind}).
Filled dots are from
Carilli et al. (\cite{car:ro}),
Chambers et al. (\cite{cham:mi}),
Djorgovski et al. (\cite{djor:spin}),
Dunlop\&Peacock (\cite{dun}),
Eales et al. (\cite{eal:raw}), 
Hammer\&LeFevre (\cite{ham:le}), 
Kristian et al. (\cite{kri:san}),
Lacy et al. (\cite{lac:mi}), 
LeFevre et al. (\cite{lefev:ham1}),
LeFevre\&Hammer (\cite{lefev:ham2}),
Lilly (\cite{lil1}),
Lilly (\cite{lil2}),
Owen\&Keel (\cite{owen:keel}),
Miley et al. (\cite{mi:ch}), 
Spinrad et al. (\cite{spin}) 
and filled stars are from the  ESO/Key-Project
(R\"{o}ttgering et al. \cite{rott:miley}
R\"{o}ttgering et al. \cite{rott:west}
van Ojik et al. \cite{van:rott}).
Open symbols are  $r$-magnitudes, which are transformed
as $R=r-0.4$.
The histogram of RC/USS sources
$R$-magnitudes is shown above.
The magnitudes are from K95b.
Present NOT-observations are concerned with $R\la24$. 
}
\label{fig2}
\end{figure}

Our high frequency
catalogue is based on  a sample of faint radio sources
originally discovered using the RATAN-600 radio telescope
in the "Kholod" ("Cold") experiment in 1980-81
(Parijskij et al. \cite{pari:bur}; Parijskij et al. \cite{pari:bur2}
Parijskij \& Korolkov \cite{par:kor}).
In the experiment, performed at 7.6 cm (3.9 GHz),
the strip around the sky at $\delta$=5$\degr \pm$ 20$\arcmin$
was surveyed with a limiting flux of about 4 mJy.
The RC catalogue resulted in  containing 1145 objects.
Within the inner strip of $\pm$ 5$\arcmin$
the completeness of the catalogue reaches 80\% at
the flux limit S$_{3.9} =7.5$ mJy and is almost
100\% at 15mJy (Parijskij et al. \cite{pari:bur}).
Such flux limits are really quite faint and allow one
to identify a large number of steep spectrum sources,
if a low frequency catalogue with
sufficiently faint flux limit is available.
The  UTRAO (Douglas et al. \cite{douglas})
is such a catalogue with a
flux limit of $\sim$ 100mJy at 365 MHz  (see Fig. ~\ref{fig1}).
The RATAN-600 catalogue (RC) provided the first sample 
which allowed one
to calculate the spectral index for practically all
UTRAO sources within the region covered by the "Kholod"
experiment (Soboleva et al. \cite{sobo:pari}).
Of the original sample of 840 sources (Parijskij et al. \cite{pari:bur}), 
491 sources matched those of the UTRAO catalogue.  
Soboleva et al. (\cite{sobo:pari}) could identify optically from
POSS (Palomar Optical Sky Survey) 240 sources at galactic 
latitude $>$ 20$\degr$.

\begin{table*}
\caption[]{RC/USS source parameters. 
The IAU name is in the first column followed by the
equatorial, then galactic coordinates and galactic extinction in R-band. 
The radio spectral index is in the seventh column, followed by
3.9 GHz flux density  and the LAS of the radio source.
The results of optical identification are in the last column.
The data have been taken from 
Kopylov et al. \cite{kop:goss1}, \cite{kop:goss2} 
and Parijskij et al. \cite{pari:goss}. }
\begin{flushleft}
\begin{center}
\begin{tabular}{lllrrrlrrr}
\noalign{\smallskip}
\hline
\noalign{\smallskip}
Name& R.A. &Decl &$l$&$b$&A$_{R}$& $\alpha^{365}_{3900}$ & S$_{3900}$ & LAS & m$_{R}$\\
    & B1950 &B1950& \degr  &\degr   & &         &mJy  & $[\arcsec]$    &   \\  
\noalign{\smallskip}
\hline
\noalign{\smallskip}
\object{J0406+0453}&04 03 48.22&4 39 49.7&186 &-33&0.41 &1.02 & 79 & 21.8  & 24.9\\
\object{J0444+0501}&04 41 38.68&4 55 55.79&192&-25&0.41 &1.09 & 69  & 10.8  &23.0 \\
\object{J0457+0452}&04 55 15.09&4 49 13.77&194&-22&0.23 &1.12 & 56  & 34    &19.4 \\
\object{J0459+0456}&04 56 25.51&4 51 30.45&194&-22&0.23 &0.95 & 76  & 63.8  &22.1 \\
\object{J0506+0508}&05 03 45.56&5 04 21.1 &195&-20&0.27 &0.88 & 70  & 0.8   &21.6 \\
\object{J0552+0451}&05 50 16.92&4 46 49.9 &201&-10&1.51 &1.18 & 65 & 1.6 &$>$25.5 \\
\object{J0743+0455}&07 40 36.54&5 03 02.88&214& 13&0.12 &1.07 & 37  & 20.5  &23.5 \\
\object{J0756+0450}&07 53 31.2 &4 47 17.1 &216& 16&0.07 &1.16 & 14 & $<$1&$>$25.0 \\
\object{J0837+0446}&08 34 51.28&4 54 51.82&221& 25&0.07 &1.0  & 54  & 3.9   &22.4 \\
\object{J0845+0444}&08 42 53.28&4 53 52.9 &222& 27&0.09 &1.14 & 135  & 4.6  &21.4 \\
\object{J0909+0445}&09 07 13.51&4 56 37.0 &225& 32&0.08 &1.0  & 64  & 1     &20.6 \\
\object{J0934+0505}&09 31 48.21&5 17 10.76&229& 38&0.06 &1.07 & 36  & 5     &24.4 \\
\object{J1031+0443}&10 28 43.01&4 58 33.53&240& 49&0.06 &1.2  & 191 & 33    &22.5 \\
\object{J1043+0443}&10 41 10.27&4 56 12.62&243& 52&0.06 &1.14 & 37  & 48    &23.0 \\
\object{J1113+0436}&11 11 24.05&4 54 20.12&252& 57&0.10 &0.98 & 52  & 29    &22.4 \\
\object{J1152+0449}&11 49 49.71&5 04 56.75&268& 63&0.07 &1.0  & 29  & 7  &$>$24.0 \\
\object{J1155+0444}&11 52 45.43&5 00 13.86&269& 63&0.07 &1.0  & 54  & 13    &18.6 \\
\object{J1219+0446}&12 17 06.94&5 04 02.84&282& 66&0.06 &1.23 & 23  & 118   &22.0 \\
\object{J1235+0435}&12 33 16.52&4 49 26.7 &292& 67&0.06 &0.98 & 45  & 7     &21.5 \\
\object{J1322+0449}&13 19 31.84&5 04 28.13&322& 66&0.06 &0.96 & 47  & 7     &20.4 \\
\object{J1333+0451}&13 30 32.35&5 07 08.5 &328& 65&0.05 &1.3  & 11  & 1     &23.4 \\
\object{J1333+0452}&13 30 54.66&5 07 21.17&328& 65&0.05 &1.4 & 16  & 54    &23.3 \\
\object{J1339+0445}&13 37 06.5 &5 10 15.85&332& 64&0.06 &1.07 & 41  & 34    &22.7 \\
\object{J1347+0441}&13 44 37.58&4 57 16.48&336& 63&0.06 &0.98 & 43  & 1.4   &23.5 \\
\object{J1429+0501}&14 26 45.73&5 14 43.41&353& 57&0.06 &0.92 & 82 &11.1 &$>$24.0\\
\object{J1436+0501}&14 34 04.66&5 15 10.8 &356& 56&0.06 &1.25 & 48  & 15    &22.9 \\
\object{J1439+0455}&14 37 15.64&5 08 38.68&357& 55&0.06 &1.15 & 40 & 17.9&$>$24.0 \\
\object{J1510+0438}&15 07 43.00&4 50 51.72&  4& 50&0.06 &0.9  & 67  & 3.4   &22.1 \\
\object{J1609+0456}&16 06 54.69&5 07 50.48& 16& 38&0.14 &1.15 & 30 & 6.3 &$>$24.5 \\
\object{J1626+0448}&16 24 21.72&4 55 33.4 & 19& 34&0.18 &1.26 & 39  & 2.4   &22.9 \\
\object{J1646+0501}&16 44 24.94&5 06 28.92& 22& 29&0.27 &0.92 & 54  & 15.7  &21.2 \\
\object{J1658+0454}&16 55 43.34&4 58 04.9 & 23& 27&0.27 &1.25 & 31&$<$0.3&$>$24.5\\
\object{J1703+0502}&17 01 01.3 &5 06 20.0 & 24& 26&0.27 &1.18 & 175 & 1.8   &23.6 \\
\object{J1720+0455}&17 17 36.0 &4 56 48.0 & 26& 22&0.27 &1.22 & 19  & $<$0.5&20.6 \\
\object{J1725+0457}&17 23 04.58&5 00 05.0 & 27& 21&0.27 &1.26 & 27  & 1  &$>$24.0 \\
\object{J1735+0454}&17 33 13.52&4 57 07.37& 28& 19&0.39 &1.0  & 30  & 4     &23.5 \\
\object{J1740+0502}&17 38 06.03&5 04 11.1 & 29& 18&0.41 &1.2  & 32  & 4     &22.5 \\
\object{J2013+0508}&20 10 54.69&5 01 24.78& 47&-15&0.46 &0.96 & 51  & 10    &21.1 \\
\object{J2036+0451}&20 34 27.46&4 39 22.7 & 50&-20&0.27 &1.02 & 75  & 56    &19.0 \\
\object{J2144+0513}&21 41 56.65&4 57 26.1 & 61&-34&0.18 &1.06 & 72  & $<$5.5&18.8 \\
\noalign{\smallskip}						       
\hline								       
\end{tabular}							       
\end{center}
\end{flushleft}
\end{table*}

\subsection{
Construction and properties of the RC/USS sample}

The present study is concerned with
sources in the range 4$^{h}<RA<22^{h}$ 
(Parijskij et al. \cite{pari:bur}).
The first RC/USS sample consists of 40 
steep spectrum ($\alpha >$ 0.9, \, $f_{\nu}\propto\nu^{-\alpha}$),
double or triple FRII sources,
and optically fainter than the POSS limit. The radio morphology
comes from observations with the VLA 
(Kopylov et al. \cite{kop:goss1}).
The largest angular size of the radio source (LAS) 
was not used as a criterion, because
only eight sources had LAS larger than 30".
The median LAS of the sample is 7$\arcsec$. 
The median 365 MHz flux density is 0.5 Jy (average 0.7 Jy) 
ranging from 0.2 Jy to 3 Jy.

Optical identifications were made from deep observations at
the 6 m telescope, down to about $m_{R}$=24.  These results
and the optical fields around the sources have been reported
by Kopylov et al. (\cite{kop:goss2}, here after K95b).  
Table 1 contains information on the basic RC/USS sample:
source name, equatorial and galactic coordinates, spectral index,
flux, LAS and $m_{R}$. 
From this list we selected
objects which are not unreasonably faint
($m_{R} <$ 24 mag) for a medium sized telescope.

Fig.~\ref{fig2} gives a representative
 m$_{R}$-$z$ Hubble diagram for radio galaxies
collected from the literature, together with the magnitude distribution
of the RC/USS objects.  
The Hubble diagram allows one to estimate
a lower limit to redshift, because of the rather sharp lower envelope,
especially above $m_{R}$=21. Where the bulk of the RC/USS galaxies 
are situated, redshift is expected to be $\ga$0.7 as shown
in Fig. 2.   
Soboleva et al. (\cite{sobo:pari}) estimated the maximum 
photometric redshifts for the RC/USS objects from 
the requirement that radio
luminosity is not higher than optical luminosity: when radio flux
is known, the minimum optical magnitude may be calculated,
hence the rough maximum $z_{ph}$, which is usually large, $>$1.

It should be mentioned that one optically
bright ($m_{R}$=19) object RC2036+0451  was measured at the
6 m telescope  to have $z$=2.95 (Pariskij et al. \cite{pari:sobo}).
Though for a quasar, this large $z$ also supports the view that
present  selection criteria lead to high average redshift.

The aim of the NOT imaging
was to study the morphology
of the RC/USS sources with high resolution
and confirm the optical identifications.
This paper is organised as follows. 
In Sect. 2  we describe our observations
and data reduction. Morphology of individual galaxies 
is discussed
in Sect. 3. The results are  summarised  in Sect. 4.

\section{Observations and reductions}

\subsection{Observations}

Optical images were obtained with the 2.56 m
Nordic Optical Telescope (NOT)
at La Palma during three observing runs in March,
May and December 1994.
Table 2 summarises the instrumentation used. 
In addition, we have
some supplementary observations from other observing runs.
I-band observations of RC1510+0438 were made with 
``Stockholm'' CCD in July 1994 and
RC2013+0508 was observed with Brocam1 in September 1994.
The complete log of observations is given in Table 3.
For each observed object it contains the filter used, 
number of separate images,
total integration time, seeing, and date.  
Calibration stars from Landolt (\cite{landolt}) were
observed several times each night at a range of air masses.

\begin{table*}
\caption[]{Instruments}
\begin{flushleft}
\begin{tabular}{lllll}
\noalign{\smallskip}
\hline
\noalign{\smallskip}
CCD & Date & Field size (pixels) & Field size ($\arcmin$)&
Pixel size ($\arcsec$) \\
\noalign{\smallskip}
\hline
\noalign{\smallskip}
Astromed &&&&\\
EEV P88200  & March 1994 & 1152x770 & 3.1x2.1& 0.163\\
IAC CCD  &&&&\\
THX31156 & May 1994 & 1024x1024 & 2.4x2.4 &
0.14\\
Brocam 1 &&&&\\
TK1024A& December 1994 & 1024x1024& 3x3 & 0.176\\
\noalign{\smallskip}
\hline
\end{tabular}
\end{flushleft}
\end{table*}

\begin{table*}
\caption[]{Journal of observations.    }
\begin{flushleft}
\begin{tabular}{cccrll|cccrll}
\noalign{\smallskip}
\hline
\noalign{\smallskip}
Object & Band & No. of & T$_{int}$ & Seeing & Date & 
Object & Band & No. of & T$_{int}$ & Seeing & Date     \\
     &      & images     & $[sec]$  & $[\arcsec]$ & 1994  & 
&     & images   & $[sec]$  & $[\arcsec]$&1994 \\
\noalign{\smallskip}
\hline
\noalign{\smallskip}
RC0444+0501 & R& 5 & 3300 & 2.0 &3.12. &RC1219+0446 & R &5 & 3000 & 0.5&18.5 \\
            & I &3 & 2700 & 2.0 & 4.12.&RC1235+0435 & V &5 & 3000 & 3 & 14.3 \\
RC0457+0452 & B &3 & 3200 & 3.2-3.9& 3.\&5.12.&    & V &3 & 2100 & 0.6 &17.5 \\
            & V &2 & 1800 & 0.7 & 2.12.& 	 & R &3 & 1800 & 3 & 14.3\\ 
            & R &4 & 3600 & 0.8 & 2.12.&	& R &3 & 1800 & 0.6&17.5\\ 
  & I &4 & 2100 & 2.0-2.9 & 4.\& 5.12. & RC1322+0449 &V&6 &3600 & 1.8 & 14.3\\
RC0459+0456 & V &2 & 1800 & 0.7& 2.12.&	       & R &3 & 1800 & 1.8 & 14.3\\
            & R &2 & 1800 & 0.6 & 2.12.& RC1333+0451 & V &1 & 600 &0.6 & 18.5\\
            & I &3 & 2700 & 2.0 & 3.12.& 	    & R &3 & 2100 & 0.5&18.5 \\
RC0506+0558 & V &2 & 1200 & 0.8& 2.12.&	RC1339+0445 &V &5 & 3000 & 1.4 & 15.3\\
            & R &6 & 1800 & 0.8 & 2.12.&	 & R &3 & 1800 & 1.4 & 15.3\\ 
            & I &2 & 1200 & 1.5 & 4.12.& RC1347+0441  & V &5 &3300 & 1.1 &16.\\
RC0743+0455 & V &3 & 1800 & 1.3 & 15.3&	           & R &3 & 2400 & 0.7 &16.5 \\
            & R &3 & 1800 & 1.1 & 15.3&	 RC1510+0438& V &3 & 2100 & 0.6 &17.5\\
            & R &1 &  600 & 0.8 & 2.12.&	& R &7 & 4100 & 0.5&17\&18.5 \\
RC0837+0446 & V &3 & 1800 & 0.9 & 14.3&	RC1609+0456 & V &1& 600 & 0.6  &18.5 \\
            & R &4 & 2400 & 1.0 & 14.3&	        & R &2 &  600 & 0.5&18.5   \\
            & I &3 & 2400 & 1.6 & 4.12.&RC1626+0448& V &2 & 1200 & 1.7 & 15.3\\
RC0845+0444 & V &1 &  600 & 1.2 & 15.3&	       & R &2 & 1200 & 1.7 & 15.3\\
            & R &1 &  600 & 1.1 & 15.3&	RC1646+0501&V&3&1800 & 1.5-2.0 & 13.3\\
            & I &2 & 2800 & 1.2 & 4.12.&     & R &3 & 1800 & 1.5-2.0 & 13.3\\  
RC0909+0445 & B &3 & 2700 & 1.8& 3.12.&	RC1703+0502 &R &4 & 2400 & 0.6 & 17.5\\
       & B &4 & 4500 & 3.0 & 5.12.&RC1720+0455& V &2 & 1200 & 1.5-2.0 & 13.3\\ 
     & V &2 & 1200 & 1.7 & 14.3        &  & V &1 &  600 &0.7   &16.5   \\      
            & V &2 & 1800 & 1.6 & 3.12.&  & R &1 &  600 & 1.5-2.0 & 13.3\\ 
            & R &2 & 1200 & 1.7 & 14.3&	   & R &2 & 1200 & 0.7 &16.5     \\ 
            & R &4 & 3600 & 2.0 & 3.12.& RC1735+0454&R&12& 3080& 0.6&17\&18.5\\
           & I &1 &  900 & 1.5 & 4.12. & RC1740+0502&V &3 & 1800 & 0.6 &16.5\\
            & I &3 & 2700 & 3.0 & 5.12.&           & R &2& 1200 & 0.6 &16.5 \\
RC1031+0443 & V &6 & 3600 &1.5-2.0&13.3&RC2013+0508 & V &2&1800&1.5&3.\&4.12.\\
      & V &3 & 1800 & 1.5 & 15.3 &         & R &1 & 900  & 1.2 & 4.12.\\
            & R &3 & 1800 & 1.6 & 13.3 &   & R &1 &  600 & 0.6 & 4.9. \\
            & R &3 & 1800 & 0.9 & 15.3 &RC2036+0451&B&2 &1800 & 1.4 & 2.12.\\
            & I &2 & 1800 & 1.4 & 4.12.& & V &2 & 1800 & 1.5 & 3.\& 4.12.\\ 
RC1043+0443 & V &5 & 3000 & 1.7 & 14.3 & & R &1 &  900 & 1.1 & 4.12.\\ 
            & R &4 & 2400 & 1.7 & 14.3 & & I &1 &  900 & 3.0 & 5.12.\\ 
RC1113+0436 & V &5 & 3000 & 1.6 & 15.3 &RC2144+0513 & B &2 & 1800&1.8 & 3.12.\\
        & R &3 & 1800 & 1.6 & 15.3 &     & V &2 & 1800 & 1.0 & 2.12.\\ 
RC1152+0449 & V &5 & 3000 & 1.0 &16.5&   & R &3 & 1920 & 1.0 & 2.12.\\  
            & R &3 & 1800 & 0.8 &16.5&   & I &2 & 1800 & 1.2 & 4.12.\\   
RC1155+0444 & V &1 &  600 & 0.6 &17.5  \\
            & R &1 &  600 & 0.6 &17.5    \\  
\noalign{\smallskip}
\hline
\end{tabular}
\end{flushleft}
\end{table*}

In this paper,
we shall restrict the discussion  to observations
made under excellent or good seeing (FWHM $\la 1\farcs$1) 
conditions,  totaling 22 objects.
We present only R-band images except for  a few cases where
the morphology has a strong wavelength dependence and the
S/N-ratio in other passbands is high enough.
All observations presented were made under
photometric conditions. 

``Blooming'' of the CCD was a serious problem with the IAC CCD.
Some objects were close to bright stars which limited
the longest possible exposure time, or a bright star had to be
moved outside the CCD.
RC1735+0454 lies close to the galactic plane, hence
the field is  crowded with bright stars. 
We could not obtain long exposures of this faint object 
and had to move it close to the edge of the CCD.
Note that the exposure time of the greyscale image is 
900 seconds and for the contour image 3080 seconds.
The bright star northeast from the centre of gravity of RC1219+0446
hampered the observations and the northern part of the radio
source was not observed.

\subsection{Reductions}

The reductions were carried out using standard IRAF routines
(bias subtraction, trimming, flat fielding).
The average bias frame was constructed for each night.
The flatfielding was made by twilight flats
obtained each evening and morning.
All the scientific frames were flattened at better than a 1\%-level.
The exposures of each object were registered in position
using several stars in the field and then averaged. The number of
reference stars varied from three up to a  dozen.

The astrometric calibration was carried out using the
APM Catalogue (Irwin et al. \cite{irwin}) whenever  possible.
For the objects near the galactic plane
the Guide Star Catalog (GSC) (Lasker et al. \cite{lasker})
was used. 
Due to the small field of view of the CCDs there
were typically only a few reference stars in the frame.
The number of stars and hence the accuracy
of the astrometry  strongly depends on
galactic latitude. We estimate the accuracy
of the astrometric calibration to be typically better than
1 second of arc. This is enough for the current study,
because the typical resolution of the radio map is
about 1$\farcs$5
and most of the radio sources are so compact that the optical
identification is straightforward.

As a check of our photometry in the March and May 1994 run
we measured comparison stars
of OJ287 (Fiorucci \& Tosti \cite{fiorucci}). 
The derived brightnesses
were consistent with each other within 0.1 magnitudes.

\begin{table*}
\caption[]{NOT imaging data. The diameter of the aperture is 
indicated in arcseconds. The magnitudes are without 
correction of galactic extinction.  Ellipticity
and position angle of resolved sources is measured
with the same aperture as the magnitudes.
The radio position angles are measured from Kopylov et al.
(\cite{kop:goss1}).
}
\begin{flushleft}
\begin{center}
\begin{tabular}{llllllcrr}
\noalign{\smallskip}
\hline
\noalign{\smallskip}
Name & Aperture & $m_{R}$ & merr&$m_{V-R}$&merr&$e$&PA$_{opt}$ & PA$_{radio}$\\
\noalign{\smallskip}
\hline
\noalign{\smallskip }
RC0457+0452 &7   &19.72 &0.01 &0.93  &0.05   &0.13    &13    &  58 \\
RC0506+0558 &3.5 &21.54 &0.03 &1.29  &0.13   &unresolved & .. & -75 \\
RC0837+0446 &6.5 &22.19 &0.08 &0.74  &0.13   &0.23    & 76   & -67 \\
RC0845+0444 &8.2 &20.75 &0.08 &1.15  &0.17   &0.25    & -78  &9  \\
RC1031+0443 &6.5 &22.32 &0.12 &1.12  &0.23   &0.57    & 88   &-36  \\
RC1152+0449 &3   &22.23 &0.07 & 0.84 &0.15   &0.11    & -33  &-15  \\
RC1155+0444 &8.4 &18.81 &0.02 & 1.09 &0.05   &0.28    & -54  &-76 \\
RC1235+0453 &4.2 &21.70 &0.06 & 0.96 &0.17   &0.26    & 43   &-50 \\
RC1347+0441 &2   &23.99 &0.19 & 0.75 &0.43   &0.32    & -29  &-49 \\
RC1510+0438 &3   &22.20 &0.04 & 1.57 &0.25   &0.08    & -79  &62 \\
RC1703+0502 &3   &23.69 &0.18 & ..   & ..    &0.49    & -86  &-82 \\
RC1720+0455 &4.2 &20.34 &0.02 & 1.15 &0.07   &unresolved & ..&point \\
RC1740+0502 &3   &22.19 &0.08 & 0.74 &0.12   &0.05  & 66  &64 \\
RC2013+0508 &5.3 &20.70 &0.04 & 0.22 &0.06   &unresolved& ..  &-55  \\
RC2036+0451 &4.2 &19.06 &0.02 & 0.35 &0.04   &unresolved & ..  &-2  \\
RC2144+0513 &3.5 &18.89 &0.02 &0.24  &0.03   &unresolved& .. &point \\
\noalign{\smallskip}
\hline
\end{tabular}
\end{center}
\end{flushleft}
\end{table*}

\section{Reduced images: Overview of the morphology}

In this section  
we give a greyscale image and a contour map
for each identified  object (Fig.~\ref{fig4}) 
The close environment and faint features can be 
studied from the greyscale image and
the confidence level of the features and light distribution
from the contour map.  
The field of view of the greyscale image is indicated in the
upper left hand corner. 
The images are slightly smoothed with a Gaussian function
($FWHM=\frac{1}{2}seeing$) in order to enhance the low 
surface brightness features and maintain the resolution
of the original images.
The centre of gravity of the radio source and
the positions of the radio lobes are indicated with a cross
and circles, respectively. The images are presented in
linear scale from 0$\sigma$ to 10$\sigma$ above
the background of the image.
In contour maps the object is in the origin and 
the numbers on both the vertical and horizontal
axes refer to distance in  arcsecond.
The contour interval is 0.5 mag arcsec$^{-2}$ and
the surface brightness of the lowest
contour is indicated in the upper right hand corner. 
The limiting surface brightness at which objects can be detected
is typically between 25 and 26 mag arcsec$^{-2}$.
For RC1510+0438 we also give V and I band images.
Uncertain identifications are presented in Fig. \ref{fig5}
and ``faint objects'' in Fig. ~\ref{fig6}.

The magnitudes are based on aperture photometry using
DAOPHOT. The size of the aperture was selected
in such a way that 1) as much light as possible was included
in the aperture while keeping the errors reasonable but 
2) the companions were excluded.
The image shapes are determined by the moments of the 
brightness distribution
using IMEXAMINE (eq. 4 in Valdes et al. \cite{valdes})
with the size of the aperture the same as in photometry.  
The estimation is vague for small 
ellipticity (e.g. RC1152+0449) or when the inner regions  
have a different position angle than the outer regions
(e.g. RC1031+0443). The results of  photometry and
 image shape analysis are given in Table 4.
The magnitudes between this work and K95b
are generally in agreement.
The differences are primarily caused by
the better seeing conditions at NOT, as compared with the 6 m-telescope
and the different size of the aperture.

\subsection{Identified objects}

In this section we give notes on individual objects (Fig. ~\ref{fig4}).
For the strongest sources alternative names are given in brackets.

\noindent
{\bf \object{RC0457+0452}}\\
This is one of the brightest objects in this  sample.
The new VLA map shows that the radio  source has FRI 
structure (an unpublished radio map).
This agrees with a high optical to
radio luminosity ratio (Parijskij et al. \cite{pari:goss}).
There is a near companion $\sim$2.5$\arcsec$ northwest from the nucleus.
In addition there are four faint companions in the southeast (5$\arcsec$).
The position angle of the galaxy is not uniform,
which is possibly due the close companions.
The inner isophotes are roughly perpendicular to the radio axis
and the outermost fuzz is roughly aligned with the radio axis. 
The same trend can be seen in both R and V-band images.
This object is possibly situated in a cluster of galaxies and
there are a few companions with similar brightness. The 
companion 20$\arcsec$ to the east has a double nucleus
($m_{R}=19.26, m_{V-R}=0.69$), hence it is 
apparently a merger, and the companion 22$\arcsec$ to the 
south has distorted morphology ($m_{R}=19.94, m_{V-R}=0.93$) .

\noindent
{\bf \object{RC0506+0508}}\\
This is a faint point-like source, even under excellent seeing
conditions.  
It might be a high redshift ($z>1, M_{R}< -23$) quasar
because
in the quasar catalogue by Veron-Cetty \& Veron (\cite{ver:ver})
there are only a few  quasars fainter than    
21 magnitudes with $z<1$.
The companions 14$\arcsec$ to the northwest have a multicomponent
structure with an extended  diffuse emission. 

\noindent
{\bf \object{RC0837+0446}}\\
The galaxy is marginally resolved and lies
in or behind a galaxy cluster.
In the lower left hand corner of the grey scale image is a
trail of a solar system object.

\noindent
{\bf \object{RC0845+0444}}\\
The optical counterpart of this radio source
coincides with the western radio component.
This object is optically extended 
and there is a 
faint extension towards the  southwest.   

\noindent
{\bf \object{RC1031+0443}}\\
(\object{4C +05.43} \& \object{PKSB1028+049})
The galaxy is near the centre of  gravity of the radio source. 
The object is extended and has a multicomponent structure. 
The strongest optical emission is aligned with the radio source,
but the position angle of the outermost isophotes is
almost perpendicular to the radio axis.

\noindent
{\bf \object{RC1152+0449}}\\
This galaxy 
has two or possibly three components.
The outer isophotes of the galaxy are box-like,
but the separate components are roughly aligned with the
radio axis. 
The faint companion 4$\arcsec$ to  the west is near to the 
western radio lobe. This blue companion may be related 
to the radio source ($m_{R}=23.0, m_{V-R}\sim$0.3).

This is the only source where the astrometry of the present work
is not consistent with the result of K95b, but coincides with 
current identification in Parijskij et al. (\cite{pari:goss}).

\noindent
{\bf \object{RC1155+0444}}\\
This is the brightest galaxy in our sample.
The galaxy is elliptical and it is clearly aligned 
with the radio source.
The two neighbouring galaxies are possibly interacting
foreground galaxies.

\noindent
{\bf \object{RC1235+0453}}\\
This faint galaxy is spatially extended in our R-band images. 
The fuzz $\sim 1 \arcsec$ northwest from the nucleus 
has a clumpy structure.
Our deep images show faint low surface brightness  
companions near the object $\sim 10 \arcsec$ to the east, north and west.
In the V-band images only the 
core of the galaxy is detected.

\noindent
{\bf \object{RC1347+0441}}\\
This galaxy is the faintest of our sample. 
The object is elongated and  
the size of the optical object is roughly the same as the
radio source. 

\noindent
{\bf \object{RC1510+0438}}\\
This is the most spectacular object in our sample,
lying in a group of galaxies and having apparently
wavelength dependent properties.
In the R- and I-band image the object is almost round
(Table 4) but in the V-band the object is weakly elongated with
the same position angle as the radio source.
The redshift estimation from BVRI colours suggest
$z\sim0.6$ (Pariskij et al. \cite{pari:sobo}).
If this is the case, the strong emission lines $[$\ion{O}{ii}$]$
3727 and $[$\ion{O}{iii}$]$ 5007 would be shifted into R and I band,
respectively, possibly causing the wavelngth dependence of morphology.
However, new colours from the 6 m-telescope
do not agree with strong line contribution.  
Another consequence of such redshift would be that
this  would be one of the faintest
(intrinsically) radio galaxies in the Hubble diagram (Fig. ~\ref{fig2}).
There are three relatively bright galaxies and one faint companion galaxy
within 5$\arcsec$ of the object. 
All the companions are bluer than the object ($m_{R-I}=1.52$).
The objects towards the east, 
C1 ($m_{R}=22.54, m_{V-R}=0.51, m_{R-I}=0.27$), 
north, C2 ($m_{R}=22.34, m_{V-R}=0.81, m_{R-I}=0.87$) 
and west,C3 ($m_{R}=23.24, m_{R-I}=0.83$) could be foreground galaxies. 
In addition there is a faint companion 1$\farcs$5 north from the 
object.

\noindent
{\bf \object{RC1703+0502}}\\
(\object{PKS B1701+051})
This is one of the strongest and most compact radio sources
of the present sample.
The optical and radio axes are clearly aligned and the sizes are
almost the same.
There are a few faint galaxies in the field, but no
close companions.  
This object is possibly located behind 
a foreground galaxy cluster although some of the
field objects might be faint galactic stars.

\noindent
{\bf \object{RC1720+0455}}\\
This object is compact in radio and optical.
This suggests that it is a QSO and
the same conclusion may be drawn from radio-optical luminosity
consideration (Parijskij et al. 1996a).
The extension towards the southeast, seen in K95b, was an artifact. 
There are a few companions close to the object.
The southern companion either has a double nucleus or a dust lane. 
The wide field image shows several faint companions,
hence this galaxy is either in a cluster of galaxies or behind one.

\noindent
{\bf \object{RC1740+0502}}\\
This source was identified by K95b 
and it is marginally resolved in the R-band image.  
In the V-band image the object has an extension to the
south west in contrast to almost round morphology in R-band
(see Table 4).

\noindent
{\bf \object{RC2013+0508}}\\
This unresolved object could possibly be a quasar.
Because of its low galactic latitude ($b\sim$-15),
the field is crowded with stars.

\noindent
{\bf \object{RC2036+0451}}\\
(\object{MRC 2034+046}).
This is the second of the two triple radio sources in this sample. 
A point source coincides with the central component fairly well.   
This is the only object with known redshift ($z$=2.95
Pariskij et al. \cite{pari:goss}).
This indicates that the
absolute magnitude of this quasar is M$_{R}\approx$-29.

\noindent
{\bf \object{RC2144+0513}}\\
This object is unresolved. 
The companion 3$\arcsec$ to the southwest is most likely a
galactic star ($m_{R}=20.90, m_{V-R}=1.1$). 
The profile of this object matches 
perfectly with the average stellar profile from the same field.

\subsection{Uncertain identifications}

\noindent
{\bf \object{RC0459+0456}}\\
(\object{MRC 0456+048})
This source has two candidates for optical identification in
K95b. Id1 ($m_{R}=22.08$) is an elongated galaxy with  roughly the same 
position angle as the radio source. This object has a 
companion to the west (Id2). This is a  marginally resolved 
point-like source, hence it might be
a quasar with a host galaxy ($m_{R}=21.12$).
FWHM of the id2 0$\farcs$66 compares with FWHM of a field star
0$\farcs$62.

\noindent
{\bf \object{RC1219+0446}}\\
This is the largest radio source in our  sample
(118$\arcsec$). The nature of the source remains unclear
and it is possible that there are indeed two independent radio sources.
If this is one source, then a possible identification would be a
faint, rather round galaxy (Id1) 5$\arcsec$ from the centre of 
radio source ($m_{R}=21.9$).
On the other hand if the southern radio lobe is an independent
object, the identification could be an unresolved 
object (Id2) 2$\arcsec$ southeast from the radio source ($m_{R}=17.88$).

\noindent
{\bf \object{RC1735+0454}}\\
The possible optical counterpart is $\sim 3 \arcsec$ to the east of 
the radio source. The galaxy has several components and it is elongated
in a north south direction. 
The identification should be confirmed by future observations.

\subsection{Faint objects}

\noindent
{\bf \object{RC0743+0455}}\\
This object is very faint and hardly visible in the 30 min. exposure. 

\noindent
{\bf \object{RC1333+0451}}\\
The radio source is compact. There is a  faint 
extended emission exactly at the position of the radio source.

\noindent
{\bf \object{RC1609+0456}}\\
New 6-m telescope measurements find an object with $m_{R}\sim25.5$ 
exactly at the position of the radio source.
Our 600 second exposures are not deep enough to detect 
this object.
The bright nearby object is unresolved and BVRI photometry by K95b
suggests it to be a star.

\section{Concluding remarks}

Excepting the quasar RC2036+0451,
the present galaxies do not have measured redshifts as yet. Hence,
it is interesting to ask whether optical
morphology
provides information on the redshift
distribution of the sample.
Other properties, for example
the Hubble diagram of the RC/USS sample, suggest (Sec. 1.3) 
that it contains   galaxies with $z\ga$0.7.

Of the 22 observed objects, three are extremely faint and  
three others are either faint or have several possible 
identifications. Of the remaining 16 objects, 5 are unresolved.
This is roughly the same fraction of point sources 
which R\"{o}ttgering et al. (\cite{rott:miley})
found, but slightly more than 
Lu et al. (\cite{lu:hof}) found from their 
``distant'' sample ($S_{1.4GHz}>35$mJy), which does not have 
radio spectral index selection criteria.
Typically RC/USS objects have a multicomponent
structure with extended emission.
This compares with
$HST$ images which have shown that about 30\%  of intermediate
redshift 3CR galaxies  have distorted morphology
(De Koff et al. \cite{dekoff}).
More distant ($z\sim1$) 3CR galaxies have typically 
multicomponent structure with diffuse extended emission 
(Best et al.  \cite{best:long}).
The ellipticity ($e$) of the current sample (11 objects) ranges from 0.05
to 0.57, with a mean of 0.25.
Taking into account the measurement errors
these values agree with studies
by Rigler et al. (\cite{rigler}) for 3C galaxies (e=0.19)
 and by R\"{o}ttgering et al. (\cite{rott:miley}) USS sample (e=0.33).

Visual inspection of the images in Fig. ~\ref{fig4} suggests 
that about half of the objects have a companion 
with comparable brightness within 10$\arcsec$.
We examined  the excess of companion galaxies 
along the radio axis suggested by R\"{o}ttgering et al.
(\cite{rott:west}). Our sample  has 7 resolved objects with   
3$\arcsec<LAS<20\arcsec$. 
Only two of these have companion galaxies (RC1152+0449 and RC1510+0438). 
RC1152+0449 has one companion almost exactly on the extension of 
the radio axis.  
RC1510+0438 has two companions, the brighter one being 
along the radio axis and the fainter one being perpendicular.
It is also interesting to note that both of these ``aligned'' 
companions are about 1 magnitude fainter than the object.
Four out of 16 objects have an apparent excess of companions
and hence may be situated in a cluster of galaxies.

We can conclude that the morphology
of the present  weak radio flux USS population 
is close to that observed by e.g HST in high-$z$ radio 
galaxies.
The results of this study make it imperative to measure
spectroscopic redshifts for the RC/USS galaxies.  
Especially, the fainter flux limit makes it very interesting to see
where in the Hubble diagram (both R and K, the latter magnitudes
still lacking) the RC-sources are found (c.f. Eales et al.
\cite{eal:ra}).
Naturally, the number of well observed RC/USS galaxies
is still small. A more conclusive discussion of morphological
features must wait until we complete the observations 
in the range $22^{h}<RA<4^{h}$ for which 6 m telescope identifications
are now available. Also,  we intend to reobserve the cases
of poor seeing in the present sample.

\begin{acknowledgements}
The authors thank the anonymous referee for useful comments. 
This work has been
partially supported by grants from the  Russian Foundation
of Basic Research  95-02-03783 and 96-02-16597
and the ``Astronomy'' programme project 2-296 and 1.2.1.2.,1.2.2.4.
Yu.V.B. acknowledges the support of the Russian 
``Integration'' project N. 578.
TP acknowledges financial support from the Wihuri foundation.
This work has been supported by the Academy of Finland (project
Cosmology in the local galaxy Universe).
This research has made use of the NASA/IPAC Extragalactic 
Database (NED) which is operated by the 
Jet Propulsion Laboratory, California
Institute of Technology, under contract with the 
National Aeronautics and Space Administration.
The National Radio Astronomy Observatory is a facility of the
National Science Foundation operated under cooperative agreement by
Associated Universities,Inc.  
\end{acknowledgements}

%
%

\begin{center}
\begin{figure*}
\caption[]{Optical R-band images of identified RC/USS-sources.
North is up and east is to the left.
The size of the greyscale image is indicated in the upper right hand corner.
The position of the radio lobes and centre of  gravity are
indicated by 1$\arcsec$ diameter circles and a cross, respectively.
The contour image is a $20\arcsec\times20\arcsec$ field around the object. 
The surface brightness of the lowest contour is indicated 
in the upper right hand corner. The offset in arcseconds relative 
to the identified object is also indicated.
In addition we give  V- and I-band contour maps for RC1510+0438 and
V-band greyscale image for RC1740+0502.
See text for comments on each object. 
}
\label{fig4}
\end{figure*}
\end{center}
\begin{center}
\bf{Fig. 3} (\it{continued})
\end{center}

\begin{center}
\bf{Fig. 3} (\it{continued})
\end{center}

\begin{center}
\bf{Fig. 3} (\it{continued})
\end{center}

\begin{center}
\bf{Fig. 3} (\it{continued})
\end{center}
\begin{center}
\bf{Fig. 3} (\it{continued})
\end{center}

\begin{figure*}
\caption[]{
Grey scale images of all the uncertain identifications
(see Sect. 5.2). 
The images are produced in the same way as for Fig. ~\ref{fig4}.  
For RC0459+0456 we also give an R-band contour map. 
}
\label{fig5}
\end{figure*}


\begin{figure*}
\caption[]{
Grey scale images of the faint objects
(see Sect. 5.3). 
}
\label{fig6}
\end{figure*}


\end{document}